\documentclass[aps,pra,twocolumn,groupedaddress,showpacs]{revtex4}
\usepackage{graphicx}
\usepackage{amssymb}
\usepackage{float}
\usepackage{amsthm}
\usepackage{epsfig}

\def\balpha{\mbox{\boldmath$\alpha$}}
\begin{document}


\title{Origin of the Low Energy Structure in Above Threshold Ionization }


\author{A.S. Titi}
\email{titi@uwindsor.ca}
\author{G.W.F. Drake}
\email{gdrake@uwindsor.ca}
\affiliation{Department of Physics, University of Windsor, Windsor, Ontario, Canada N9B 3P4 }



\begin{abstract}
We present an {\it ab initio} analytic theory to account for both the very
low energy structure (VLES) [C. Y. Wu {\it et al.}, Phys.\ Rev.\ Lett.\ {\bf
109}, 043001 (2012); W. Quan {\it et al.},\ Phys.\ Rev.\ Lett.\ {\bf 103},
093001 (2009)], and the low energy structure (LES) [W. Quan {\it et al.}\
Phys.\ Rev.\ Lett.\ {\bf 103}, 093001 (2009); C.I. Blaga {\it et al.}, Nat.\
Phys.\ {\bf 5}, 335 2009)] of above threshold ionization. The origin of both
VLES and LES lies in a forward scattering mechanism by the Coulomb potential.
We parameterize the $S$ matrix in terms of $\balpha$, which is the displacement of the the classical
motion of an electron in the laser field. When $\balpha=0$, the $S$ matrix is singular, which we attribute
to be forward Coulomb scattering without absorption of light quanta.
By devising a regularization scheme, the resulting $S$ matrix is non-singular when $\balpha=0$, and the origins of
VLES and LES are revealed. We attribute VLES to multiple forward scattering
of near-threshold electrons by the Coulomb potential, with no absorption of
light quanta, signifying the role of the Coulomb threshold effect. We
attribute LES to be due to the combined role of the Coulomb threshold effect
and  rescattering in the forward direction by the Coulomb potential with the
absorption of light quanta. A comparison of theory with experiment confirms
these conclusions. Further more, recently Dura {\it et al.} [Sci.\ Rep.\ {\bf 3}, 2675 (2013)] reported the detection of slow electrons at near zero momentum, at 1.3 meV, which is much below the VLES, almost at threshold. Our theoretical formulation gives rise to slow electrons at near zero momentum and at threshold. In addition, for circularly polarized fields, it conserves the angular momentum in the ionization process which necessitate the disappearance of the VLES, LES and the slow electrons near threshold.
\end{abstract}


\maketitle

Above threshold ionization (ATI) is a phenomenon characteristic of the
interaction of high-intensity lasers with atoms in which the atom absorbs
many more photons than the minimum number required to produce ionization.
The general oscillatory features of the photoelectron spectrum are well
understood within the strong-field approximation (SFA) ( Zeroth-order term in
the expansion of the S-matrix in terms of the atomic potential) of Keldysh-Faisal-Reiss
(KFR) \cite{Keldysh,Faisal,Reiss1}, but recently, unexpected spike-like
structures \cite{Blaga1,Quan,Wu,Catoire} have been reported in the low energy
region around 1 eV. This threshold structure is in striking contrast to the
predictions of the KFR theory. Furthermore, during the preparation of this paper, Dura {\it et al} \cite{Dura} reported another surprise, the detection of slow electrons at 1.3 meV, almost at threshold.

The structures seem to be a universal feature of ATI in 
atoms and molecules. They consist of two humps: the first hump, called VLES,
lies below 1 eV and weakly depends on the laser wavelength \cite{Wu,Quan}.
The second hump, called LES, is characterized by a peak energy 1 eV $<
E_{{k}} < $3.7 eV which is typically observed using midinfrared lasers, and
extends to higher energies (2--20 eV) where the ATI spectra merge with the
predictions of KFR theory \cite{Blaga1,Quan,Catoire}.  Beyond this, there is
a plateau around $2U_{{p}}$ due to direct electrons, and a higher energy
plateau around $10U_{{p}}$ due to rescattered electrons
\cite{Corkum,Lohr,Milosevic1,Milosevic2,Bao}, where $U_{{p}}$ is the
ponderomotive energy. The goal of this paper is to present a full quantum
mechanical theory of the unique LES and VLES structures.  They appear in the
tunneling regime where the Kyldysh parameter
$\gamma=\sqrt{E_{\rm{B}}/2U_{{p}}}<<1$, with $E_{\rm{B}}$ being the atomic
ionization energy.  In this region, one would normally expect the SFA to be
valid.

Various theoretical investigations have been carried out to understand the
origin of the low energy structures. Numerical solution of the time dependent
Schr\"{o}dinger equation (TDSE) \cite{Blaga1,Wu} provide quantitative agreement but
little physical insight. The semiclassical model \cite{Liu,Yan} has been applied to explore the LES and
revealed the essential role of the Coulomb potential in its
production via forward scattering mechanism. Recently,
 Guo {\it et al.}\ presented an {\it ad hoc heuristic} quantum mechanical calculation
demonstrating that the origin of the LES lies in the Coulomb
interaction. However, the role of the Coulomb interaction in the production
of the VLES is not well understood. Guo {\it et al.} attempted to account for both the VLES and the LES
in terms of rescattering , but their calculations failed to display the VLES. In this paper, we present an {\it ab initio }  analytical quantum mechanical formulation that simultaneously accounts for both the LES and the VLES. Contrary to the speculation of Guo {\it al.,}\ we show that the VLES is due to Coulomb threshold effects via forward scattering (with no absorbtion of light quanta), in accordance with Wigner threshold law  for Coulomb attraction \cite{Wigner}, and the LES is due to  Forward rescattering (with absorbtion of light quanta). The {\it ab initio} analytical approach does provide a transparent ideas of the process; thus making the understanding of VLES and LES  complete. In addition, it does give rise to slow electrons near zero momentum and at threshold, in accordance with Wigner threshold law, consistent with the findings of Dura {\it et al.,}\ which we attribute to  multiple Coulomb forward scattering. Furthermore, the Coulomb interaction is fundamental in physics, and so the Coulomb singularity and its regularization  is of broad interest.

We start with the exact expression for the time reversed transition amplitude
from a ground state $\phi_{\rm{i}}$ to a final continuum state
$\Psi_{\rm{f}}^{-}$ \cite{Reiss1} (unless  specified atomic units
are used throughout)
\begin{equation}
 (S-1)_{\rm{fi}} = -\imath\int_{-\infty}^{\infty}dt\langle\Psi_{\rm{f}}^{-}|V_{\rm{L}}\phi_{\rm{i}}\rangle
 \end{equation}
The final state $\Psi_{\rm{f}}^{-}$ is a solution to the TDSE for an atomic electron interacting with a laser field
\begin{equation}
(\imath\partial_{{t}}-H_{0}-V_{\rm{L}}-V_{\rm{A}})\Psi_{\rm{f}}^{-}({\bf r},t)=0
\end{equation}
Here $V_{\rm{A}}$ is the atomic Coulomb potential, and
$ V_{\rm{L}}=\frac{1}{c}{\bf A}(t)\cdot\hat{P}+\frac{A(t)^{2}}{2c^{2}}
$
is the atom-laser interaction Hamiltonian where $\hat{P}=-\imath{\bf\nabla}$ is the momentum operator and ${\bf A}(t)$ is the vector potential of the laser field.

To first order in  $V_{\rm{A}}$, Eq.\ (1) reads
\begin{equation}
(S-1)_{\rm{fi}}  \approx  S^{({0})}_{\rm{fi}} + S^{({1})}_{\rm{fi}}
\end{equation}
 $S^{({0})}_{\rm{fi}}$ is the  KFR direct electron term
\begin{equation}
S^{({0})}_{\rm{fi}} = -\imath \int_{-\infty}^{\infty}dt\, \langle \Psi^{(\rm{v})}_{{\bf k}}(t)\mid V_{\rm{L}}(t)\phi_{\rm{i}}(t)\rangle
\end{equation}
and $S^{({1})}_{\rm{fi}}$ is the rescattered electron term
\begin{equation}
S^{({1})}_{\rm{fi}} = -\imath \int_{-\infty}^{\infty}dt \int_{-\infty}^{t}dt^{\prime}\, \langle \Psi^{(\rm{v})}_{{\bf k}}(t)\mid V_{\rm{A}}G^{(+)}_{\rm{L}} V_{\rm{L}}(t^{\prime})\phi_{\rm{i}}(t^{\prime})\rangle
\end{equation}
where $ \Psi^{(\rm{v})}_{{\bf k}}(t)$ is the Volkov wave function and $
G^{(+)}_{\rm{L}}(t,t^{\prime})$ is the retarded Volkov propagator
\cite{Volkov}. Analytical evaluation of $S^{({1})}_{\rm{fi}} $ gives
\begin{eqnarray}
S^{({1})}_{\rm{fi}} & =  & 2\pi\imath \sum_{n}\delta (E_{k}+E_{\rm{B}}+U_{{p}}-n\omega)\,\nonumber\\
 &\times&\int d{\bf q}\, \langle {\bf q}\mid \phi_{\rm{i}}({\bf r})\rangle\, \langle {\bf k}\mid V_{\rm{A}}\mid {\bf q}\rangle\nonumber \\
 & \times & \sum_{m} \frac{U_{{p}}-m\omega}{E_{q}+E_{\rm{B}}+U_{{p}}-m\omega-\imath \eta}
 \, f_{m}\, g_{n-m}
\end{eqnarray}
where $\eta$ is an infinitesimal  parameter, the matrix elements $\langle \,\mid \, \rangle$ are Fourier
transforms, and the functions $f_{m}({\bf q})$, $g_{n-m}({\bf q},{\bf k})$
are generalized and  ordinary Bessel functions respectively. For a
Coulomb potential with effective charge $Z$, $\langle
{\bf k}\mid V_{\rm{A}}\mid {\bf q}\rangle = -\frac{Z}{2\pi^{2}|{\bf k}-{\bf
q}|^{2}}$. As $\eta \rightarrow 0^{{+}}$,  the
integrand in Eq.\ (6) is divergent for $m=n$ at ${\bf q}={\bf k}$. (see supplementary material \cite{Supp}). When $n\neq m$, it is nonsingular. Thus we split
$S^{({1})}_{\rm{fi}}$ into a regular part $S^{{(1)}}_{\rm{r}}$ and irregular
(singular) part $S^{{(1)}}_{\rm{ir}}$ so that
\begin{equation}
S^{({1})}_{\rm{fi}} = S^{({1})}_{\rm{r}} + S^{{(1)}}_{\rm{ir}}
\end{equation}
where $S^{({1})}_{\rm{r}}$ is given by Eq.\ (6), but with the term $m=n$
excluded from the sum over $m$, and $S^{{(1)}}_{\rm{ir}}$ corresponds to the
$m=n$ term in Eq.\ (6).

$S^{({1})}_{\rm{ir}}$, represents scattering by the atomic core
from intermediate continuum states with  momentum ${\bf q}$  to final
continuum states with  momentum ${\bf k}$  with no exchange of
extra photons. This represents  {\em forward scattering}, ${\bf q}={\bf k}$, without changing the kinetic
energy of ionized electrons. This term does only contribute to the
near threshold low energy direct electrons via {\em forward scattering} by
the Coulomb potential, which results in a spike in the near
threshold low energy electrons. This is the origin of VLES.

$S^{(1)}_{\rm{r}}$ is not only relevant to the high energy electrons of the
ATI spectrum via {\em backward scattering } but also to the near
threshold low energy electrons via {\em forward scattering } with
the absorption of extra photons. Careful inspection of the expression for $S^{(1)}_{\rm{r}}$ given
by Eq.\ (6) ($m=n$ term is excluded), reveals that {\em forward scattering}
with the exchange of one photon  is significant whenever $m=n\pm 1$;
that is, when $E_{{q}}=E_{{k}}\pm \omega$, and ${\bf q}\parallel {\bf k}$.
Scattering in the forward direction (${\bf q}\parallel {\bf k}$), then the
matrix element $\langle{\bf k}\mid V_{\rm{A}}\mid{\bf q}\rangle\propto
\frac{1}{E_{{k}}+E_{{q}}-2\sqrt{E_{{k}}E_{{q}}}}$. Now, the LES lies in the
energy range $0.037 < E_{{k}} <0.136$ (in a.u.) and for midinfrared laser
wavelengths frequencies $\omega < .031$, which makes
$\frac{w}{E_{{k}}} < 1$ in the energy range of LES. Setting
$E_{{k}}=E_{{q}}\pm \omega$ and  using $\frac{\omega}{E_{{k}}}<1$, then
$\langle{\bf k}\mid V_{\rm{A}}\mid{\bf q}\rangle\sim\frac{1}{2E_{{k}}\mp
\omega-2E_{{k}}(1\mp
\frac{\omega}{2E_{{k}}}-\frac{\omega^{2}}{8E_{{k}}^2}+...)}\sim\frac{1}{\omega^{2}/E_{{k}}}>>1$.
The smaller $\omega$, the larger is its value. This forward scattering would be negligible had we employed a screened
Coulomb potential. Thus we are inclined to
attribute the origin of LES to  forward scattering  by the 
Coulomb potential with the exchange of photons (forward rescattering).
Precise quantitative calculations will confirm this conclusion.

Employing the Henneberger transformation \cite{Henneberger}
\begin{equation}
\Psi^{(-)}_{\rm{f}} = e^{-\imath \int^{t} d\tau\, V_{\rm{L}}(\tau)}\, \Phi^{(-)}_{\rm{f}}
\end{equation}
then  Eq.\ (2), now reads
\begin{equation}
\left(\imath \frac{\partial}{\partial t} - H_{0} - V_{\rm{A}}({\bf r}+\balpha)\right)\Phi^{(-)}_{\rm{f}} = 0
\end{equation}
where $\balpha=\frac{1}{c}\int_{-\infty}^{t} d\tau {\bf A(\tau)}$. To first order in $V_{\rm{A}}({\bf r}+\balpha)$, we have
\begin{equation}
\mid\Phi^{(-)}_{\rm{f}}(t)\rangle \approx \mid\Phi^{(0)}_{\rm{f}}(t)\rangle + \mid \Phi^{(1)}_{\rm{f}}(t)\rangle
\end{equation}
where $\mid\Phi^{(0)}_{\rm{f}}(t)\rangle=\mid\chi_{_{{\bf k}}}(t)\rangle = \,\mid{\bf k}\rangle\,e^{-\imath E_{k}t} $ is a plane wave which gives the KFR term $S^{({0})}_{\rm{fi}}$ given in Eq.\ (4) and
\begin{equation}
\mid \Phi^{(1)}_{\rm{f}}(t)\rangle  =  \int_{t}^{\infty} dt^{\prime}\, G^{(-)}_{{0}}(t,t^{\prime})\, V_{\rm{A}}({\bf r^{\prime}}+\balpha(t^{\prime}))\, \mid\chi_{_{{\bf k}}}(t^{\prime})\rangle
\end{equation}
which gives the rescattering term $S^{({1})}_{\rm{fi}}$ given in Eqs.\ (5-6).
Here $G^{(-)}_{{0}}(t,t^{\prime})$ is the advanced free particle propagator.

Now setting $\balpha=0$ (i.e.\ when the electron is in the vicinity of the
atomic core) in Eq.\ (11) and using Eq.\ (8) then the resulting $S$-matrix is
exactly the singularity in  $m=n$ term in Eq.\ (6); i.e., $S^{{(1)}}_{\rm{ir}}$ (see supplementary material \cite{Supp}).
Therefore, the singularity in the $S$-matrix is identified to be
due to a single forward scattering of near threshold electrons by the Coulomb
potential. If we denote the singular component of
$\mid\Phi^{(1)}_{\rm{f}}\rangle $ by $\mid\Phi^{(1)}_{\rm{ir}}\rangle$, obtained from
Eq. (11) by setting $\balpha=0$, then according to Botero and Macek
\cite{Botero}, $\mid\Phi^{(1)}_{\rm{ir}}\rangle$ can
be replaced by the nonsingular component
$\frac{\partial\mid\Psi^{(-)}_{\rm{A}}\rangle}{\partial\lambda}|_{{{\lambda=0}}}$,
where $\mid\Psi_{\rm{A}}^{(-)}\rangle$  are the Coulomb scattering states
given by
\begin{equation}
\mid\Psi^{(-)}_{\rm{A}}\rangle = \mid\chi_{_{{\bf k}}}\rangle e^{-\imath\pi a/2}\,\Gamma(1+a)\, _{1}F_{1}(-a,1,-\imath(kr+{\bf k}\cdot{\bf r}))
\end{equation}
 with $a=\imath\lambda Z/k$, and $\lambda$ is a small perturbation parameter
scaling the Coulomb potential. {\em This component represents a single forward scattering by the Coulomb
center of threshold electrons without changing their energy and hence the emergence of the VLES}. To
evaluate the $S$-matrix due to this component, we consider
the Coulomb scattering state to be an approximate eigenstate of $e^{-\imath
\int^{t} d\tau\, V_{\rm{L}}(\tau)}$ \cite{Jain} and recognize that the
arising space integral is a Nordsieck-type integral \cite{Nordsieck} to
obtain
\begin{eqnarray}
& &-\imath\int_{-\infty}^{\infty}dt\,\langle\,e^{-\imath \int^{t} d\tau\, V_{\rm{L}}(\tau)}\,\frac{\partial\Psi^{(-)}_{\rm{A}}}{\partial\lambda}|_{{{\lambda=0}}}|V_{\rm{L}}\phi_{\rm{i}}\rangle \nonumber \\
&=& -\imath\frac{Z}{k}[-\imath\pi/2+\gamma^{'}+\ln{(\frac{Z+\imath k}{Z-\imath k})}+\imath\frac{k}{Z}]\cdot S^{({0})}_{\rm{fi}}
\end{eqnarray}
where $\gamma^{'}$ is Euler's constant. It is clear from Eq.\ (13) that near
threshold electrons suffer the greatest single forward scattering by the
Coulomb potential. The singularity in Eq.\ (13) at $k=0$  gives rise to $1/k$ singularity in the ionization rates. As $k\rightarrow0$, the Coulomb scattering state (see Eq.\ (12))
$\Psi^{(-)}_{\rm{A}}\approx \frac{1}{2\pi}\sqrt{\frac{\lambda Z}{k}}
J_{0}(2\sqrt{(2\lambda Zr)}\cos\frac{\theta}{2})$, $J_{0}(x)$ being the
Bessel function. $S$-matrix with $\frac{1}{\sqrt{k}}$ singularity, gives a non-singular finite ionizations rates at threshold. Thus, the inclusion of Coulomb potential to all orders with multiple forward scattering is required.

Let us define $W({\bf r},\balpha)=V_{\rm{A}}({\bf r}+\balpha)-V_{\rm{A}}({\bf
r})$. Then Eq.\ (9) reads
\begin{equation}
\left(\imath \frac{\partial}{\partial t} - H_{0} - V_{\rm{A}}({\bf r}) - W({\bf r},\balpha)\right)\Phi^{(-)}_{\rm{f}} = 0
\end{equation}
The scattering states $\Phi^{(-)}_{\rm{f}}$ are given by
\begin{equation}
\mid\Phi^{(-)}_{\rm{f}}\rangle\approx\mid\Psi_{\rm{A}}^{(-)}\rangle +  \int_{t}^{\infty}dt'\,G_{\rm{A}}^{(-)}(t,t')
W({\bf r},\balpha)\mid\Psi_{\rm{A}}^{(-)}\rangle
\end{equation}
where $G_{\rm{A}}^{(-)}(t,t')$ is the Coulomb Green function. The Coulomb scattering states in Eq.\ (15) are given by Eq.\ (12) but with $a=\imath Z/k$. Replacing the advanced Coulomb propagator with the free particle one, and  using Eqs.\ (8) and (1) and defining $\tilde\Psi^{(-)}_{\rm{A}}(t) = e^{-\imath \int^{t} d\tau\, V_{\rm{L}}(\tau)}\,\Psi^{(-)}_{\rm{A}}(t)$ we obtain
\begin{widetext}
\begin{eqnarray}
 (S-1)_{\rm{fi}}  \approx  -\imath\int_{-\infty}^{\infty}dt\langle\tilde\Psi^{(-)}_{\rm{A}}|V_{\rm{L}}\phi_{\rm{i}}\rangle +\imath \int_{-\infty}^{\infty}dt \int_{-\infty}^{t}dt^{\prime}\, \langle \tilde\Psi^{(-)}_{\rm{A}}(t)\mid
 W({\bf r},-\balpha(t))G^{(+)}_{\rm{L}} V_{\rm{L}}(t^{\prime})\phi_{\rm{i}}(t^{\prime})\rangle
\end{eqnarray}
\end{widetext}
where we utilized
$e^{-\imath\int^{t}d\tau\,V_{\rm{L}}(\tau)}\,G^{(+)}_{{0}}\,e^{\imath\int^{t}d\tau\,V_{\rm{L}}(\tau)}
= G^{(+)}_{\rm{L}} $ and
$e^{-\imath\int^{t}d\tau\,V_{\rm{L}}(\tau)}\,W(\balpha)\,e^{\imath\int^{t}d\tau\,V_{\rm{L}}(\tau)}
=- W(-\balpha)$. The resulting $S$-matrix given in Eq.\ (16) is nonsingular, and includes the Coulomb interaction to all orders in the final state wave function.
The first term $S^{({0})}_{\rm{fi}}$ on the left is the direct electron term
which includes multiple forward Coulomb scattering without changing
the photoelectron energy (Coulomb Threshold effect). This term gives rise to slow electrons at threshold and to the VLES (see Fig.\ 1). The second term
$S^{({1})}_{\rm{fi}}$ is the nonsingular rescattered electron term which gives rise to the LES (see Fig.\ 1). When
$\balpha=0$, $W({\bf r},\balpha)=V_{\rm{A}}({\bf r}+\balpha)-V_{\rm{A}}({\bf
r})=0$ thus assuring the removal of the singularity arising when $\balpha=0$.
Note that for $r>>\alpha$, $W({\bf
r},\balpha)\approx\frac{-\balpha\cdot\hat{{\bf r}}}{r^{2}}$, which is just a
short range potential.
\begin{figure}[htb]
\centering
\begin{tabular}{ccc}
\hspace*{-.45cm}\includegraphics[width=175pt]{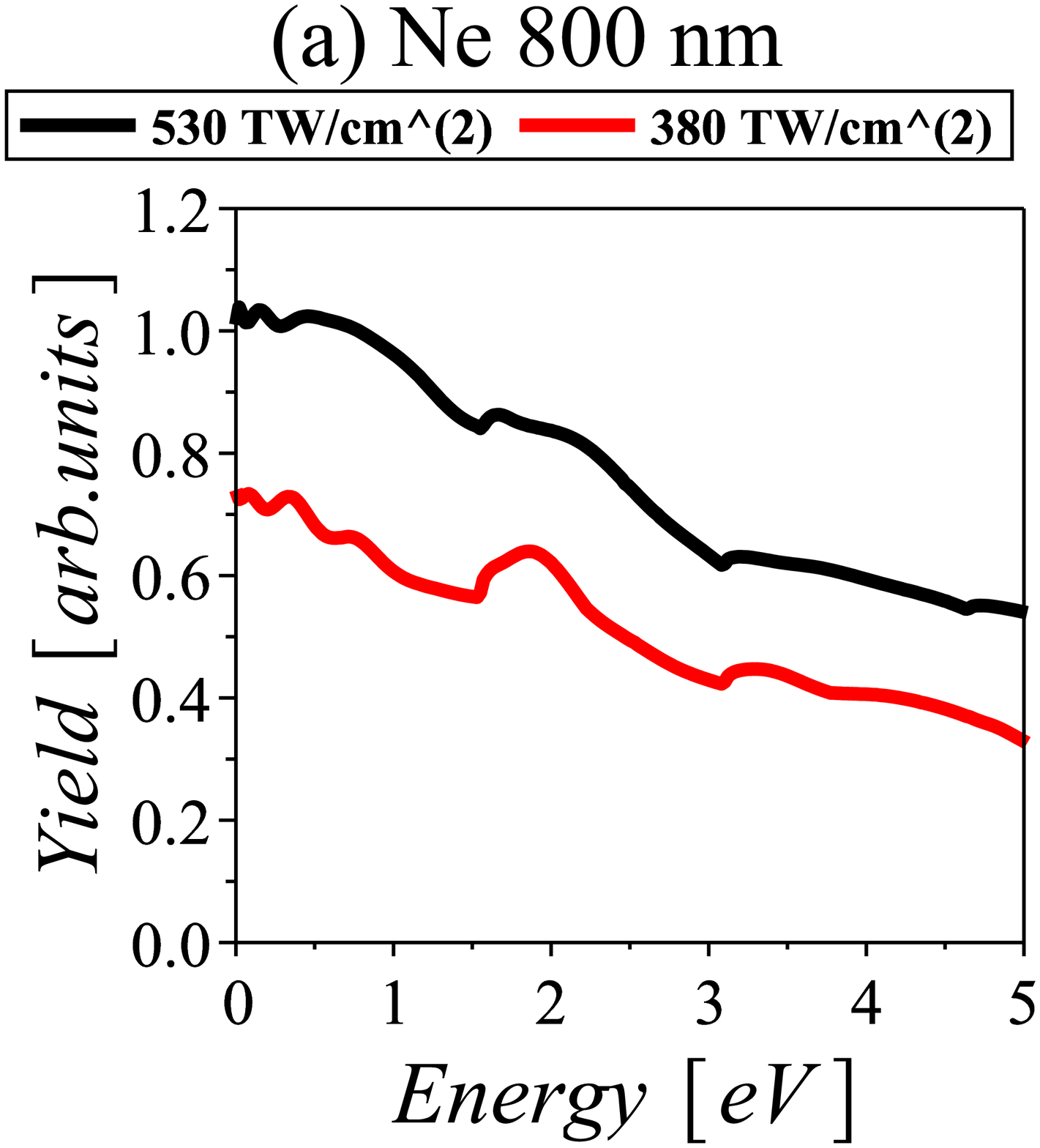}&\hspace{-3.25cm}
\includegraphics[width=175pt]{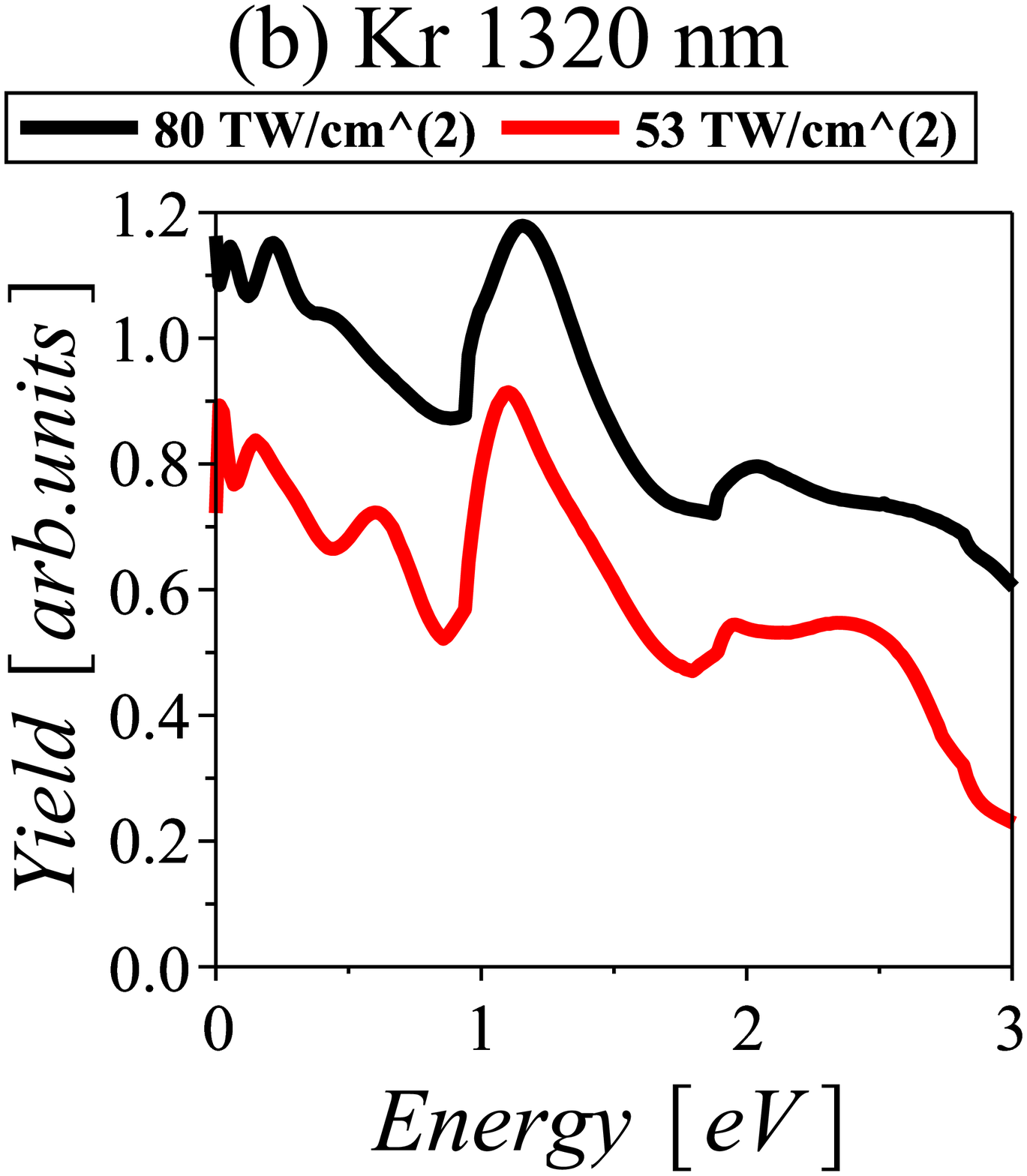}&\hspace{-3.25cm}
\includegraphics[width=175pt]{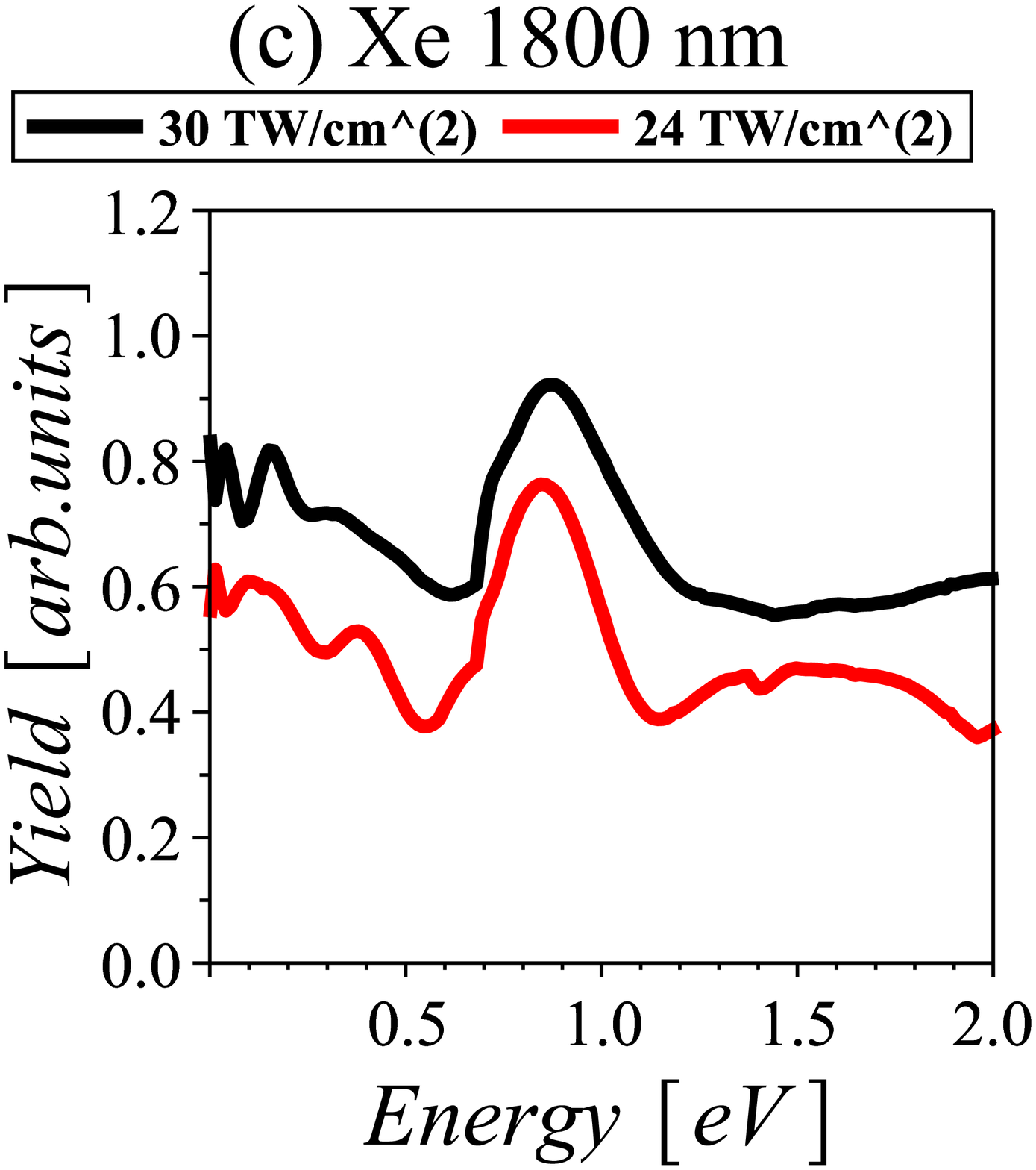}\\ [-4.0ex]
\hspace*{-.45cm}\includegraphics[width=175pt]{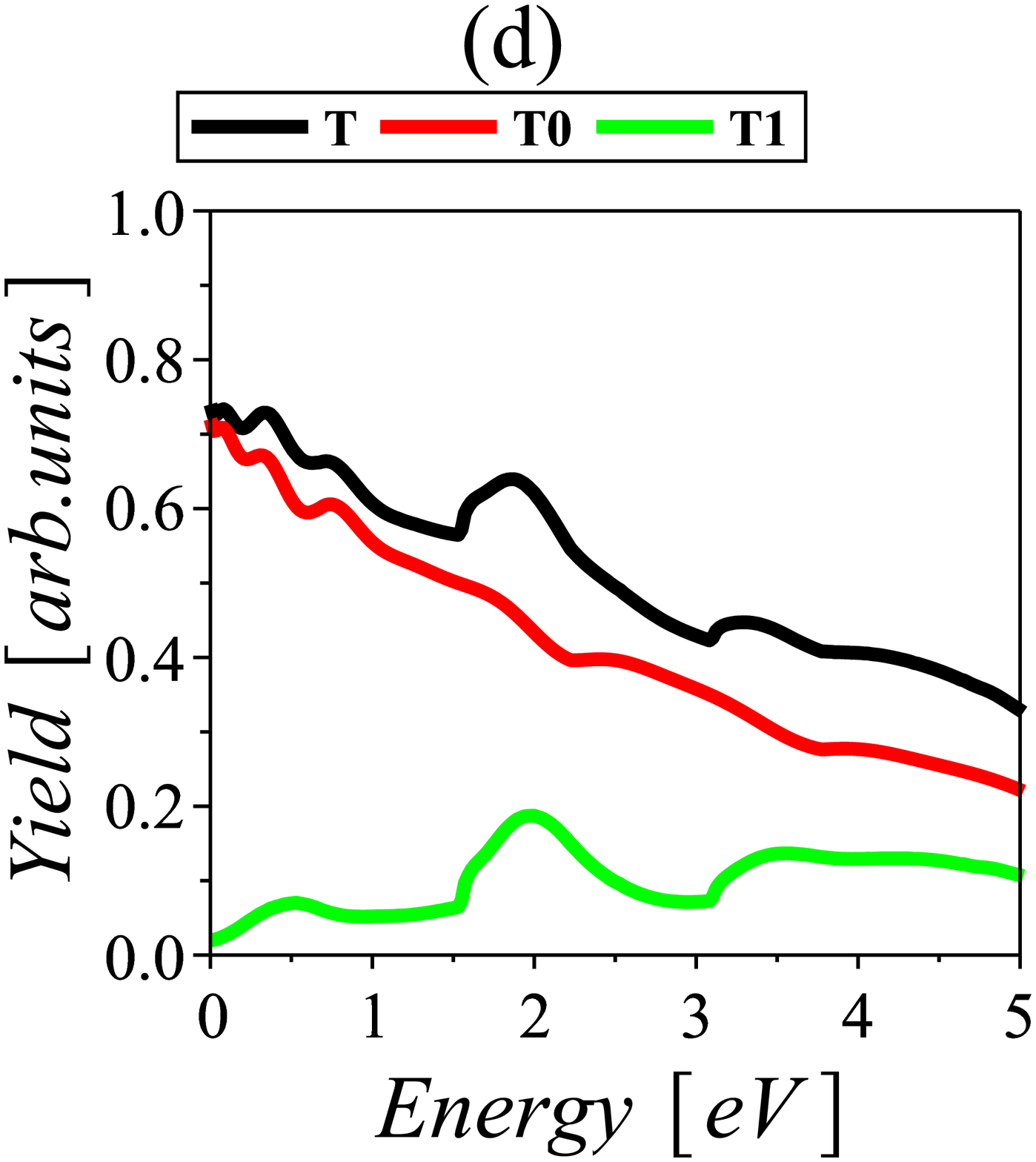}& \hspace{-3.25cm}
\includegraphics[width=175pt]{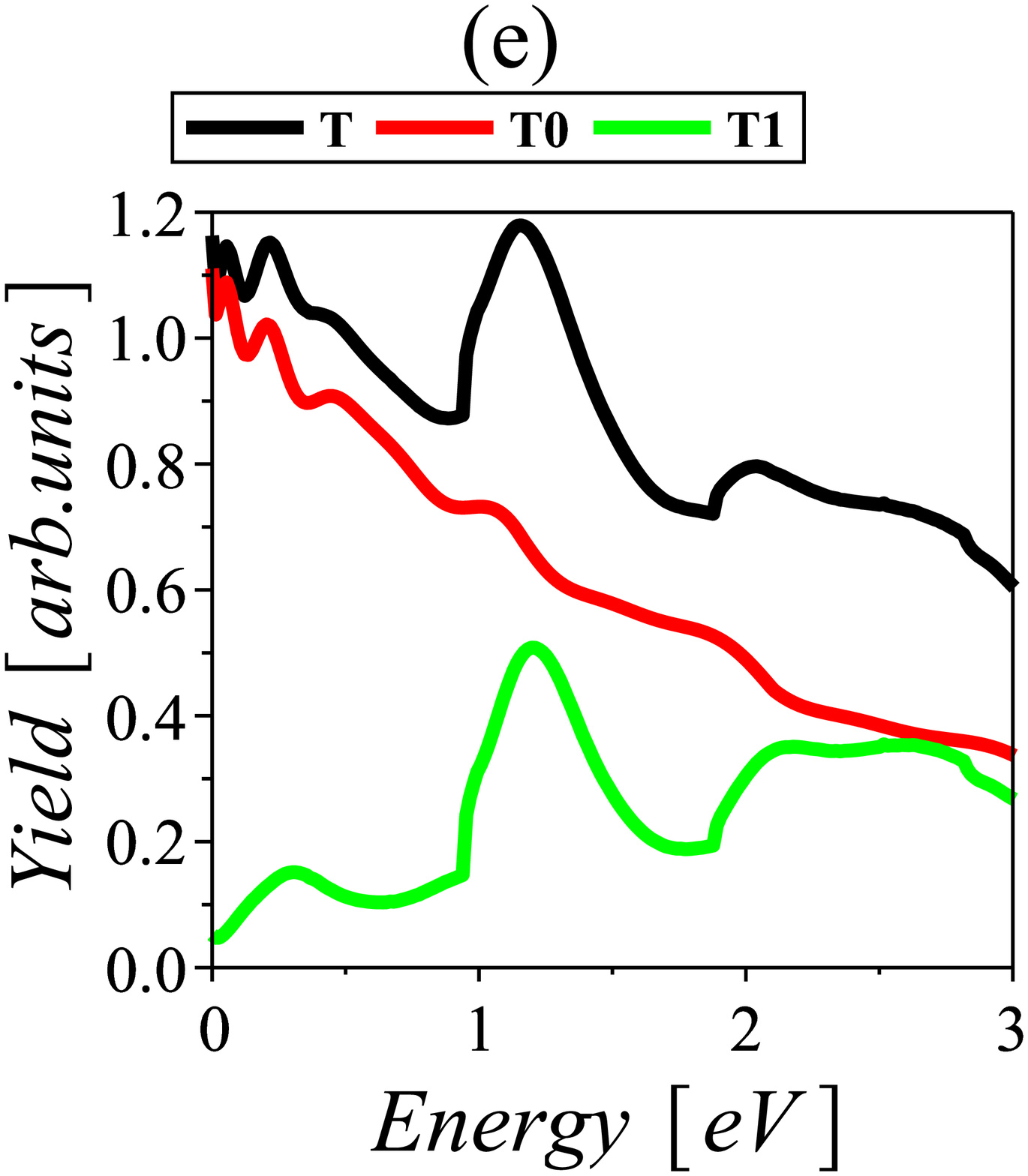}&\hspace{-3.25cm}
\includegraphics[width=175pt]{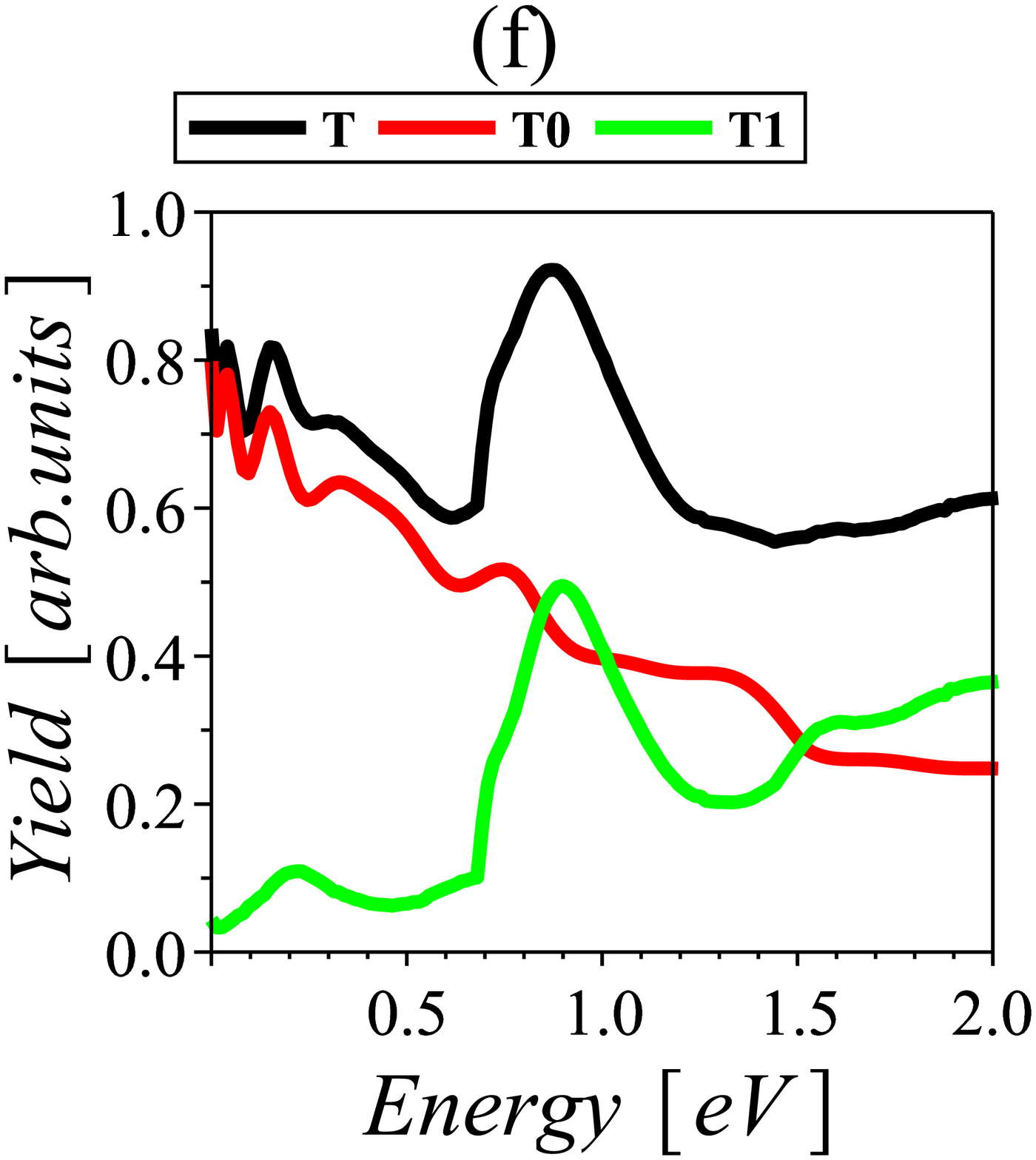} \\ [-4.0ex]
\multicolumn{3}{c}{\includegraphics[width=250pt]{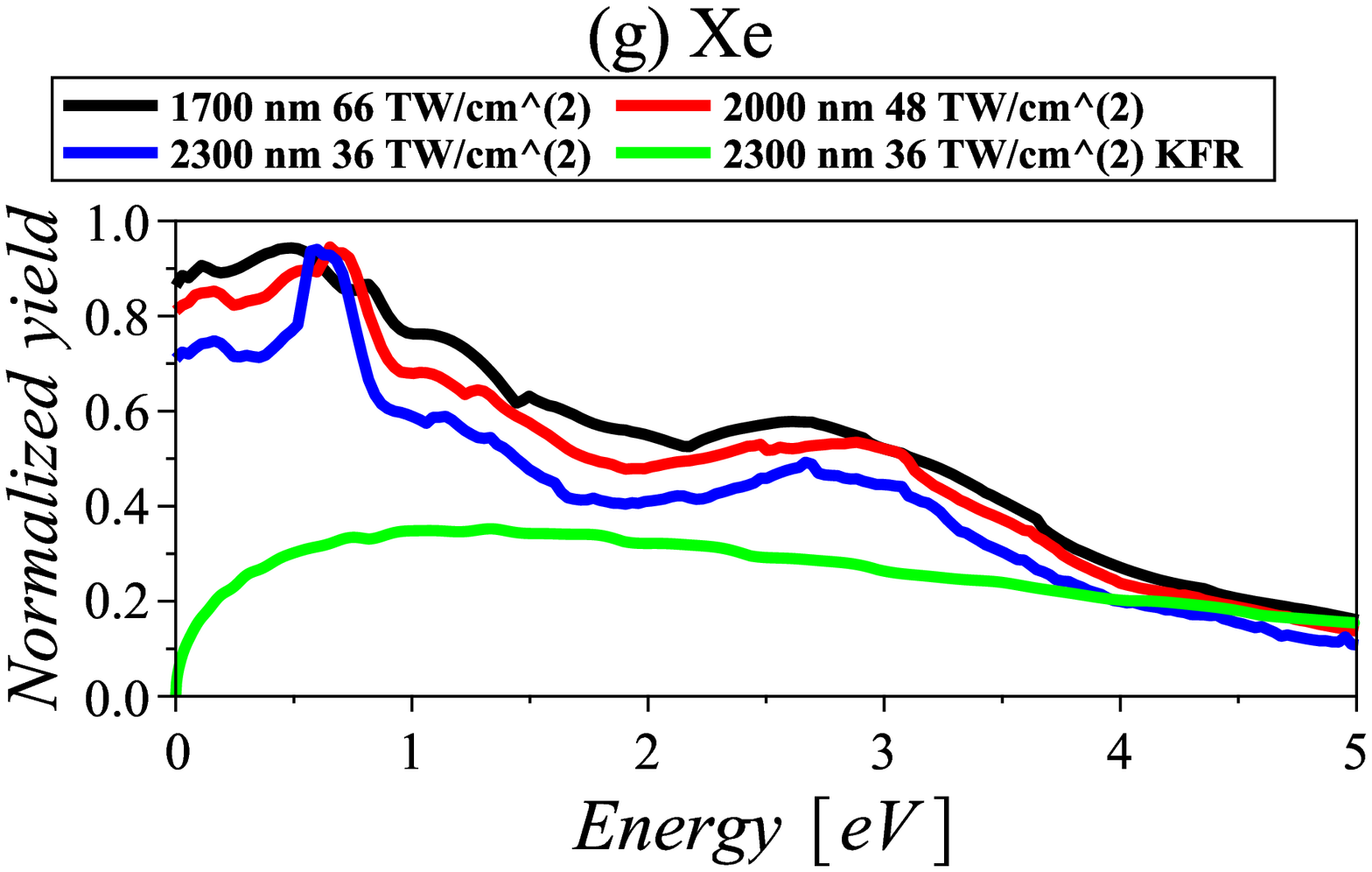}}
\end{tabular} \vspace*{-1.0cm}
\caption{(color online) Low energy photoelectron energy spectra, along polarization direction, of Ne at 800 nm [(a),(d)], Kr at 1320 nm [(b),(e)], and Xe at 1800 nm [(c),(f))]. (g) is the low energy photoelectron spectra of Xe at different intensities and wavelengths corresponding to a constant ponderomotive potential $U_{{p}}\approx18$ eV. Green curve in (g) is KFR theory prediction.}
\end{figure}

Analytical evaluation of $S^{({0})}_{\rm{fi}}$ and $S^{({1})}_{\rm{fi}}$,
given in Eq.\ (16) is nontrivial. In a previous publication we presented an
analytical evaluation of $S^{({0})}_{\rm{fi}}$ \cite{Titi1}. Full inclusion
of the Coulomb attraction in the the final state upholds the conservation of
angular momentum in the ionization process \cite{Titi2}, which is important
for circularly polarized lasers. For linearly polarized lasers,
when considering electrons ionized along polarization direction, the resulting $S^{({0})}_{\rm{fi}}$ is
the KFR term multiplied by the factor $e^{-\imath
a/2}\,\Gamma(1+a)\,e^{\imath a^{*}/2}\,\Gamma^{*}(1+a)=\frac{2\pi \imath
a}{e^{2\pi\imath a}-1}; a=\frac{\imath Z}{k}$. This would be obtained if we
replace the Coulomb scattering states $\Psi^{(-)}_{\rm{A}}$ with
$\sqrt{\frac{2\pi \imath a}{e^{2\pi\imath a}-1}}\mid\chi_{_{{\bf
k}}}\rangle$. This replacement is carried out in Eq.\ (16).

Writing
$(S-1)_{\rm{fi}}\approx-2\pi\imath\sum_{n=n_{0}}^{\infty}\delta(n\omega-E_{{k}}-E_{\rm{B}}-U_{{p}})\,T_{\rm{fi}}(n)$,
$T_{\rm{fi}}\approx T^{(0)}_{\rm{fi}}+T^{(1)}_{\rm{fi}}$ where
$T^{(0)}_{\rm{fi}}$, and $T^{(1)}_{\rm{fi}}$ are the $T$ matrices associated
with  $S^{({0})}_{\rm{fi}}$, and  $S^{({1})}_{\rm{fi}}$ respectively. According to Eq.\ (16), the squared absolute value $|T_{\rm{fi}}|^{2}$ , diverges as $1/k$
as  $k\rightarrow0$. As a result the differential ionization rate for the
absorption of $n$ photons with momentum ${\bf k}$ along the polarization
direction $\bar{\omega}_{\rm{fi}}(n)\approx2\pi
k|T^{(0)}_{\rm{fi}}+T^{(1)}_{\rm{fi}}|^{2} $ is non-vanishing. Unlike
the KFR,  the differential ionization rate $\bar\omega_{\rm{fi}}(n)$ remains
finite at $E_{{k}}=0$, in accordance with the Wigner threshold law for
Coulomb attraction \cite{Wigner}. In the vicinity of threshold it is considerably larger than that predicted by the KFR theory (see Fig.\ 1(g)), thus giving rise to slow electrons at near zero momentum, consistent with the findings of Dura {\it el al}.

\begin{figure}[htb]
\centering
\begin{tabular}{ccc}
\hspace*{-.45cm}\includegraphics[width=195pt]{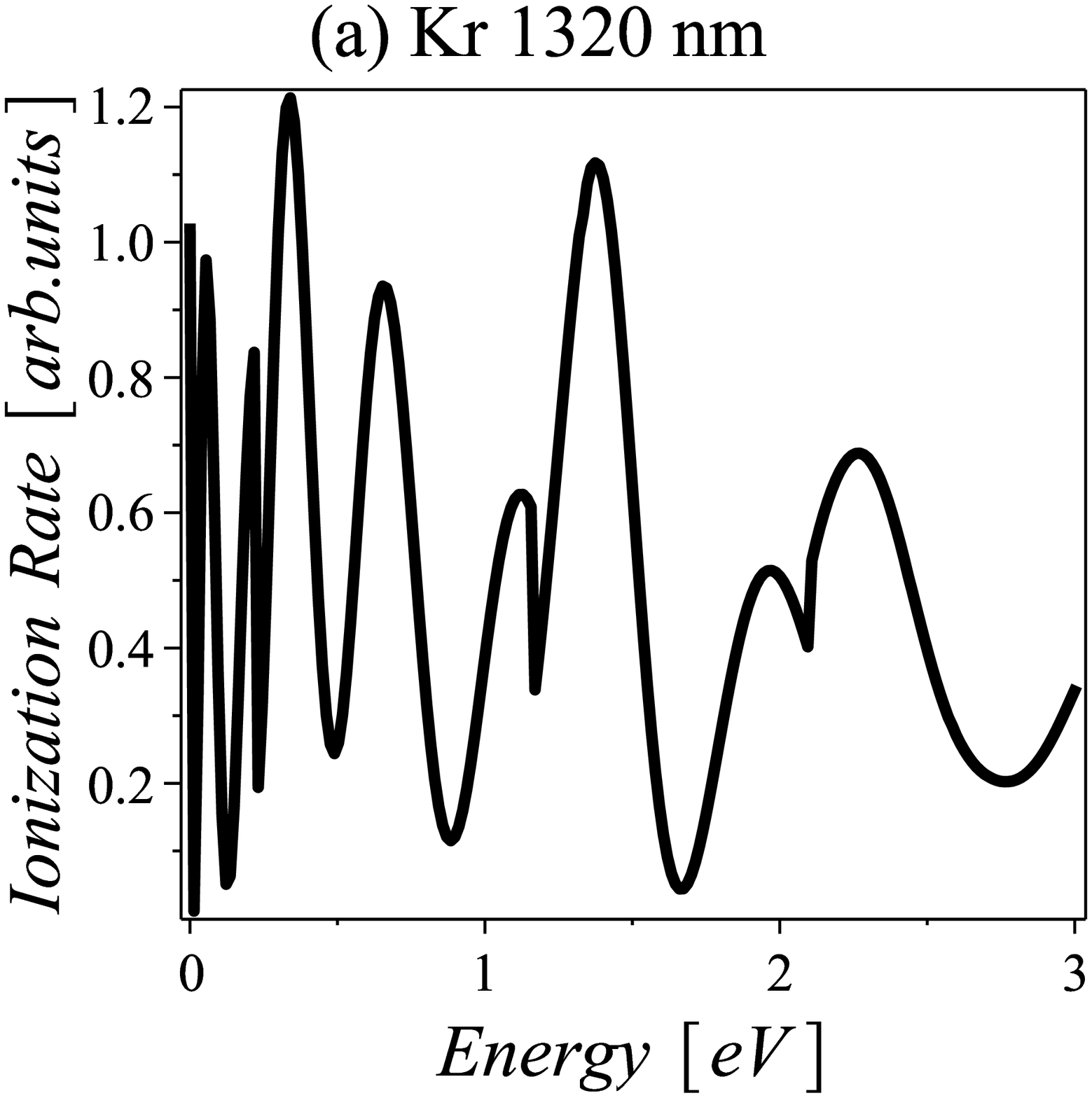}&\hspace{-2.75cm}
\includegraphics[width=195pt]{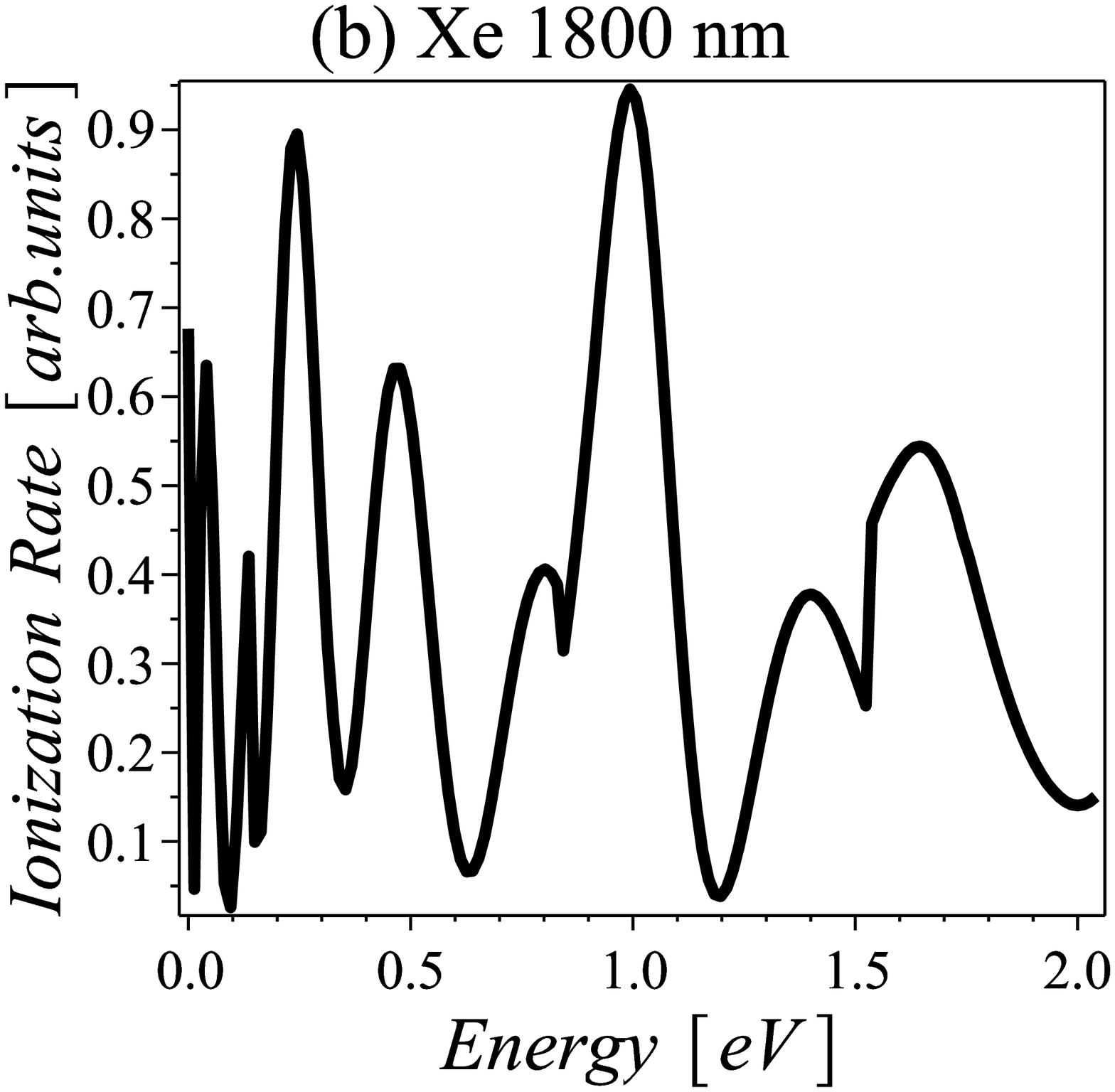}&\hspace{-3.25cm}
\end{tabular} \vspace*{-.0cm}
\caption{(color online) Non-focally averaged low energy photoelectron energy spectra, along polarization direction, of Kr at 1320 nm  and intensity 80 $TW/cm^{2}$ (a), and Xe at 1800 nm and intensity 30 $TW/cm^{2}$ (b).}
\end{figure}

In Fig.\ 1(a--f), we present the results of the theoretical calculations for
the focally averaged low energy photoelectron spectra, along the polarization
direction, of Ne [(a),(d)], Kr [(b),(e)], and Xe [(c),(f)] for the laser
parameters used in the experiment of Wu {\it et al.}\ \cite{Wu} (a Gaussian
envelope is assumed for the pulse shape). We used hydrogen-like wave functions and scattering states.
The theoretical calculations, Fig.\
1(a--c), clearly indicate a VLES, all below 1 eV---at 0.5 eV for Ne which
slightly shifts with intensity, and at 0.25 eV for both Kr and Xe,
independent of intensity. Furthermore, a LES is shown, located at energies
greater than corresponding laser frequency $\omega$ ($\ge$ 1 eV), which becomes noticeable
in the long wavelength spectra $(\lambda > 1$ $\mu$m) and shifts with
intensity. These results are in excellent agreement with the experiment of Wu
{\it et al.}\ \cite{Wu}. To reveal the origin of VLES and LES, we present in
Fig.\ 1(d--f) the yield due to $T^{(0)}_{\rm{fi}}$, and $T^{(1)}_{\rm{fi}}$ respectively for Ne at intensity 380 TW/cm$^{2}$, Kr at
intensity 80 TW/cm$^{2}$, and Xe at intensity 30 TW/cm$^{2}$. Fig.\ 1(d--f), indicates that the VLES is due to
$T^{(0)}_{\rm{fi}}$ representing multiple forward scattering with no
absorption of light quanta, and the LES
to be due to $T^{(1)}_{\rm{fi}}$ representing forward multiple scattering
with the absorption of light quanta; i.e., forward rescattering. Furthermore,
$T^{(1)}_{\rm{fi}}$ is negligible until the electron energy becomes equal to
the laser frequency $\omega$ ($\omega=1.55$ eV, 0.94 eV, and 0.69 eV for 800
nm, 1320 nm, and 1800 nm respectively), and that the LES is noticeable at
longer wavelengths and becomes prominent at $\lambda > 1.5$ $\mu$m. Thus we conclude that the VLES is located at a position below the laser
frequency $\omega$ and the LES lies beyond $\omega$. In Fig.\ 1(g), we
present the theoretical results for Xe at different intensities and
wavelengths corresponding to a constant ponderomotive energy $U_{{p}}\approx
18$ eV representing the laser parameters used in the experiments of Blaga
{\it et al.}\ \cite{Blaga1}, and Guo {\it et al.}\ \cite{Guo} as well as the
KFR theory prediction for Xe at 2300 nm and intensity 36 TW/cm$^{2}$. The results indicate the invariance of the LES and the VLES with respect to the laser
parameters if $U_{{p}}$ is held constant ( VLES is shown to be located at 0.25 eV, which wasn't reported in refs.\
\cite{Blaga1} and \cite{Guo}). To exclude the possibility that the above results are due to focal averaging,
we present in Fig.\ 2 the non-focally averaged  low energy spectra of Kr at 1320 nm and intensity 80 TW/cm$^{2}$ and Xe at 1800 nm and intensity 30 TW/cm$^{2}$. It clearly indicates a VLES, below 0.5 eV, at 0.25 eV for both Kr and Xe and a LES located at energies greater than the corresponding laser frequency $\omega$ ($\ge$ 1 eV).

In conclusion, we present an  {\it ab initio\/} analytical
theory to account for both the VLES and LES in ATI and the role of the Coulomb potential in their production.
By regularizing the $S$-matrix, their origin is revealed. We attribute the VLES
to multiple forward scattering of the near-threshold electrons by the Coulomb
potential, with no absorption of extra light quanta, signifying a Coulomb
threshold effect, and the LES to be due to the combined role of Coulomb
threshold effects and forward Coulomb rescattering. The emergence of slow electrons at threshold and near zero momentum,
which is observed by Dura {\it et al.,} is in accordance of Wigner threshold law \cite{Wigner}. The theoretical results under the same laser
parameters as used in the experiments of Wu {\it et al.}\ \cite{Wu}, Blaga
{\it et al.}\ \cite{Blaga1}, and Guo {\it et al.}\ \cite{Guo} confirms these conclusions. Contrary to Ref.\ \cite{Guo}, the
formulation presented here produced a rescattering term which is over-all
smaller than the direct term. Finally, embedded in the analytical formulation, a conservation of angular momentum in the ionization process. The disappearance of the low energy structures when circularly or elliptically polarized light is used \cite{Blaga1, Dura}, and the emergence of the LES at higher energy when elliptically light is used \cite{Dura}, is a consequence of the conservation of momentum in the ionization process. This will be thoroughly investigated in a future publication.

Research support by the Natural Sciences and Engineering Research Council of
Canada, SHARCNET, and Compute/Calcu Canada is gratefully acknowledged.

\bibliography{basename of .bib file}

\end{document}



\title{Supplementary Material to : \\
 Origin of the Low Energy Structure in Above Threshold Ionization }


\author{A.S. Titi}
\email{titi@uwindsor.ca}
\author{G.W.F. Drake}
\email{gdrake@uwindsor.ca}
\affiliation{Department of Physics, University of Windsor, Windsor, Ontario, Canada N9B 3P4 }



\pacs{42.50.Hz, 32.80.Fb, 32.80.Rm}

\maketitle

\section{Demonestration of the Logarithmic Divergence of the S Matrix in the Forward Direction}
\label{Demonestration of the Logarithmic Divergence of the S Matrix in the Forward Directionn}
In the text we state that the $m=n$ term in Eq.\ (6) in the text is singular in the forward direction and gives rise to the singular part $S_{\rm{ir}}^{(1)}$. Setting $m=n$ in Eq.\ (6) and using the energy conserving delta function, then we have
\begin{eqnarray}
S^{({1})}_{\rm{ir}} & =  & 2\pi\imath \sum_{n=n_{{0}}}^{\infty}\delta (E_{k}+E_{\rm{B}}+U_{{p}}-n\omega)\,\nonumber\\
 &\times&\int d{\bf q}\, \langle {\bf q}\mid \phi_{\rm{i}}\rangle\, \langle {\bf k}\mid V_{\rm{A}}\mid {\bf q}\rangle\nonumber \\
 & \times &  \frac{U_{{p}}-n\omega}{E_{q}-E_{\rm{k}}-\imath \eta}
 \, f_{n}\, g_{0}
\end{eqnarray}
The product $\langle {\bf q}\mid \phi_{\rm{i}}\rangle\,f_{n}({\bf q})\, g_{0}({\bf q,k})$ is analytical function in ${\bf q}$, while the matrix element $\langle {\bf k}\mid V_{\rm{A}}\mid {\bf q}\rangle\ \sim |{\bf k-q}|^{-2}$ is singular when ${\bf q=k}$. Now writing
\begin{equation}
\langle {\bf q}\mid \phi_{\rm{i}}\rangle\,f_{n}\, g_{0}=\int d{\bf r}\, F({\bf r,k})\, e^{\imath{\bf q\cdot r}}
\end{equation}
we obtain
\begin{eqnarray}
S^{({1})}_{\rm{ir}} & \sim & \sum_{n=n_{{0}}}^{\infty}\delta (E_{k}+E_{\rm{B}}+U_{{p}}-n\omega)\,(U_{\rm{p}}-n\omega)\nonumber\\
 &\times&\int d{\bf r}\, F({\bf r,k}) \int d{\bf q}\, \frac{e^{\imath{\bf q\cdot r}}}{|{\bf k-q}|^{2}(E_{q}-E_{\rm{k}}-\imath \eta)}
\end{eqnarray}
The integral over ${\bf q}$ is divergent in the forward direction and the divergence is logarithmic, reminiscent of the long range Coulomb interaction. Since $E_{\rm{k}}=k^{2}/2$ and $E_{\rm{q}}=q^{2}/2$ and using Feynmann two denominator integral formula \cite{Mathew}
\begin{equation}
\frac{1}{CD}=\int_{0}^{1}\, \frac{dx}{[Dx+C(1-x)]^{2}}
\end{equation}
so that
\begin{eqnarray}
S^{({1})}_{\rm{ir}} & \sim & \sum_{n=n_{{0}}}^{\infty}\delta (E_{k}+E_{\rm{B}}+U_{{p}}-n\omega)\,(U_{\rm{p}}-n\omega)\nonumber\\
 &\times&\int d{\bf r}\, F({\bf r,k})\int_{0}^{1} dx\, e^{\imath(1-x){\bf k\cdot r}}\,\int d{\bf p}\,\frac{e^{\imath{\bf p\cdot r}}}{[p^{2}-\beta^{2}]^{2}} \nonumber\\
 & &
\end{eqnarray}
with
\begin{eqnarray}
{\bf p} & = & {\bf q}-(1-x){\bf k} \nonumber\\
\beta^{2} & = & x(xk^{2}+2\imath\eta) \nonumber
\end{eqnarray}
By choosing the z-axis along ${\bf p}$ then the integral over the angles is straightforward and the remaining integral over $p$ is evaluated using the residue theorem of complex variable theory so that
\begin{eqnarray}
S^{({1})}_{\rm{ir}} & \sim & \sum_{n=n_{{0}}}^{\infty}\delta (E_{k}+E_{\rm{B}}+U_{{p}}-n\omega)\,(U_{\rm{p}}-n\omega)\nonumber\\
 &\times&\int d{\bf r}\, F({\bf r,k})\,e^{\imath{\bf k\cdot r}}\,\int_{0}^{1} dx\, \frac{e^{[-\imath{\bf k\cdot r}x+\imath\sqrt{x^{2}k^{2}+2\imath x\eta }r]}}{\sqrt{x^{2}k^{2}+2\imath x\eta}} \nonumber\\
 & &
\end{eqnarray}
$S^{({1})}_{\rm{ir}}$ as given by the above equation has a logarithmic divergence in the limit $\eta\rightarrow0$. To see this we write
\begin{eqnarray}
S^{({1})}_{\rm{ir}} & \sim & \sum_{n=n_{{0}}}^{\infty}\delta (E_{k}+E_{\rm{B}}+U_{{p}}-n\omega)\,(U_{\rm{p}}-n\omega)\nonumber\\
 &\times&\int d{\bf r}\, F({\bf r,k})\,e^{\imath{\bf k\cdot r}}\left\{\int_{0}^{1}dx\,\frac{e^{[-\imath{\bf k\cdot r}x+\imath\sqrt{x^{2}k^{2}+2\imath x\eta}\,r]}-1}{\sqrt{x^{2}k^{2}+2\imath x\eta}}\right.\nonumber \\
 & & \left. +\int_{0}^{1}\frac{dx}{\sqrt{x^{2}k^{2}+2\imath x\eta}}\right\}
\end{eqnarray}
The second integral on the right hand side of Eq.\ (7) is easily evaluated to be equal to $-\frac{1}{k}\ln{\frac{\imath\eta}{2k^{2}}}$.

Now we set $\imath(kr-{\bf k\cdot r})x=y$ so that for small $\eta$ we have
\begin{eqnarray}
S^{({1})}_{\rm{ir}} & \sim & \frac{1}{k}\sum_{n=n_{{0}}}^{\infty}\delta (E_{k}+E_{\rm{B}}+U_{{p}}-n\omega)\,(U_{\rm{p}}-n\omega)\nonumber\\
 &\times&\int d{\bf r}\, F({\bf r,k})\,e^{\imath{\bf k\cdot r}}\left\{\int_{0}^{\imath(kr-{\bf k\cdot r})}dy\,\frac{e^{y}-1}{y}\right.\nonumber \\
 & & \left. -\ln{\frac{\imath\eta}{2k^{2}}}\right\}
\end{eqnarray}
The integral over $y$ is identified as an integral representation of the exponential integral function $\rm{Ei}(z)$ \cite{Abramowitz} and so we have
\begin{eqnarray}
S^{({1})}_{\rm{ir}} & \sim & \frac{1}{k}\sum_{n=n_{{0}}}^{\infty}\delta (E_{k}+E_{\rm{B}}+U_{{p}}-n\omega)\,(U_{\rm{p}}-n\omega)\nonumber\\
 &\times&\int d{\bf r}\, F({\bf r,k})\,e^{\imath{\bf k\cdot r}}\left\{\rm{Ei}[\imath(kr-{\bf k}\cdot {\bf r})]\right. \nonumber \\
 & & \left.-\ln[\imath(kr-{\bf k\cdot r})]-\gamma'-\ln(\frac{\imath\eta}{2k^{2}})\right\}
\end{eqnarray}
where $\gamma'$ is Euler's constant. In the limit $\eta\rightarrow0$, $S^{({1})}_{\rm{ir}} $ is logarithmically divergent. Since the divergence is an additive not a multiplicative then it can be subtracted in a regularization scheme and can be removed.
\section{Isolation of the Singularity}
\label{Isolation of the Singularity}
We will show now that setting $\balpha=0$ in Eq.\ (11) in the text gives the singularity in  $m=n$ term in Eq.\ (6); i.e, in $S^{({1})}_{\rm{ir}}$. First we need to state that $g_{0}({\bf q,k})$ appearing in Eq.\ (1) in the supplementary material is a Bessel function of order 0, $J_{0}(x)$, such that when ${\bf q=k}$, its argument becomes equal to 0. Thus $g_{0}({\bf q,k})=1$ when ${\bf q=k}$. Taking this into account we can write the expression for $S^{(\rm{1})}_{\rm{ir}}$  given in Eq.\ (1) in the supplementary material as (without any loss of generality, if the value of $g_{0}({\bf q,k})\neq 1$ when ${\bf q=k}$, the final result is the same)
\begin{eqnarray}
S^{({1})}_{\rm{ir}}&  = &  2\pi\imath \sum_{n=n_{{0}}}^{\infty}\delta (E_{k}+E_{\rm{B}}+U_{{p}}-n\omega)\,
 \int d{\bf q}\, \langle {\bf q}\mid \phi_{\rm{i}}\rangle\, \langle {\bf k}\mid V_{\rm{A}}\mid {\bf q}\rangle
  \frac{U_{{p}}-n\omega}{E_{q}-E_{\rm{k}}-\imath \eta}
 \, f_{n}\, (g_{0}-1) \nonumber \\
 &+& 2\pi\imath \sum_{n=n_{{0}}}^{\infty}\delta (E_{k}+E_{\rm{B}}+U_{{p}}-n\omega)\,
 \int d{\bf q}\, \langle {\bf q}\mid \phi_{\rm{i}}\rangle\, \langle {\bf k}\mid V_{\rm{A}}\mid {\bf q}\rangle
  \frac{U_{{p}}-n\omega}{E_{q}-E_{\rm{k}}-\imath \eta}
 \, f_{n}
\end{eqnarray}
The dominant contribution to $S^{({1})}_{\rm{ir}}$ comes from the second part in Eq.\ (10) which is very large when ${\bf q \approx k}$ and singular when ${\bf q=k}$. Now
\begin{equation}
\frac{1}{E_{\rm{q}}-E_{\rm{k}}-\imath\eta}=P(\frac{1}{E_{\rm{q}}-E_{\rm{k}}})+\imath\pi\delta(E_{\rm{q}}-E_{\rm{k}})
\end{equation}
The principal value does not contribute and the $\delta$ function causes $q=k$ which makes the argument of the Bessel function $g_{0}$ to be very close to zero, which makes $(g_{0}-1)\approx 0$. Thus the contribution from the first term is not only negligible compared to the singular second term but also very negligible on its own.

Setting $\balpha=0$  in Eq.\ (11) in the text ($V_{\rm{A}}({\bf r}+\balpha)=V_{\rm{A}}({\bf r})$) and using Eqs.\ (8) and (1) in the text, then the resulting S matrix  can be written as
\begin{equation}
(S-1)_{\rm{fi}} = -\imath \int_{-\infty}^{\infty}dt \int_{-\infty}^{t}dt^{\prime}\, \langle \chi_{_{{\bf k}}}(t)\mid V_{\rm{A}}({\bf r})G^{(+)}_{0}(t,t') \,e^{\imath\int V_{\rm{L}}(\tau) d\tau}\,V_{\rm{L}}(t^{\prime})\phi_{\rm{i}}(t^{\prime})\rangle
\end{equation}
Now we write the free particle propagator as
\begin{equation}
G^{(+)}_{0}(t,t^{\prime}) = -\imath \Theta (t-t^{\prime})\int d\vec{q}\, \mid \chi_{_{{\bf k}}}(t)\rangle \langle \chi_{_{{\bf k}}}(t^{\prime})\mid e^{-\eta(t-t^{\prime})}
\end{equation}
where, $\eta \rightarrow 0^{+}$ is implied by the outgoing boundary conditions. Thus the resulting S matrix now reads
\begin{eqnarray}
S^{(\rm{1})}_{\rm{fi}}& = & -\imath \int d\vec{q}\int_{-\infty}^{\infty}dt\, \langle \chi_{_{{\bf k}}}(t)\mid V_{\rm{A}} \mid \chi_{_{{\bf q}}}(t)\rangle \nonumber \\
&\times & (-\imath)\int_{-\infty}^{t}dt^{\prime}\, \langle\,e^{-\imath\int V_{\rm{L}}(\tau) d\tau}\, \chi_{_{{\bf q}}}(t^{\prime})\mid V_{\rm{L}}(t^{\prime})\phi_{\rm{i}}(t^{\prime})\rangle e^{-\eta(t-t^{\prime})} \nonumber \\
& &
\end{eqnarray}
Let us denote $\Gamma(t)$ to be
\begin{equation}
\Gamma(t)=(-\imath)\int_{-\infty}^{t}dt^{\prime}\, \langle\,e^{-\imath\int V_{\rm{L}}(\tau) d\tau}\, \chi_{_{{\bf q}}}(t^{\prime})\mid V_{\rm{L}}(t^{\prime})\phi_{\rm{i}}(t^{\prime})\rangle e^{-\eta(t-t^{\prime})}
\end{equation}
the plane wave $\mid\chi_{_{{\bf q}}}\rangle=\mid{\bf q}\rangle\,e^{-\imath E_{\rm{q}}t}$ is an eigenvector of $V_{\rm{L}}$ with $V_{\rm{L}}\mid\chi_{_{{\bf q}}}\rangle=V_{\rm{L}}({\bf q})\mid\chi_{_{{\bf q}}}\rangle$. Writing
\begin{eqnarray}
\mid\phi_{\rm{i}}(t')\rangle &=& \mid\phi_{\rm{i}}\rangle\, e^{\imath E_{\rm{B}}t'} \nonumber \\
e^{-\imath\int V_{\rm{L}}(\tau) d\tau}\, \mid\chi_{_{{\bf q}}}(t^{\prime})\rangle&=& e^{-\imath S({\bf q},t')}\,\mid{\bf q}\rangle \nonumber
\end{eqnarray}
 where $S({\bf q},t)=\frac{1}{2}\int_{-\infty}^{t'}d\tau\,({\bf q}+\frac{1}{c}{\bf A}(\tau))^{2}$ is the semiclassical action of a free electron inside laser field. Proceeding and carrying out integration by parts over $t'$ we obtain
 \begin{eqnarray}
\Gamma & = & -\langle \vec{q}\mid \phi_{\rm{i}}\rangle \left\{ e^{\imath (S(\vec{q},t)+ E_{\rm{B}}t)}(-\imath) (E_{\rm{q}}+E_{\rm{B}}-\imath\eta)\right.\nonumber \\
& & \left.\times\int_{-\infty}^{t}dt^{\prime}\, e^{\imath( S(q,t^{\prime})-E_{\rm{q}}t^{\prime}-U_{\rm{p}}t^{\prime})}\, e^{\imath(E_{\rm{q}}+E_{\rm{B}}+U_{\rm{p}})t^{\prime}}\, e^{-\eta (t-t^{\prime})}\right\} \nonumber \\
& &
\end{eqnarray}
$S(\vec{q},t^{\prime})-E_{\rm{q}}t^{\prime}-U_{\rm{p}}t^{\prime}$ is periodic in $t^{\prime}$ with period equals $\frac{2\pi}{w}$ so we write
\begin{equation}
e^{\imath(S(\vec{q},t^{\prime})-E_{\rm{q}}t^{\prime}-U_{\rm{p}}t^{\prime})} = \sum_{n=-\infty}^{\infty} f_{n}\, e^{-\imath nwt^{\prime}}
\end{equation}
where $f_{n}$ are the Fourier components. Substituting Eq.\ (17) into Eq.\ (16) we obtain
\begin{eqnarray}
\Gamma & = &-\langle \vec{q}\mid \phi_{\rm{i}}\rangle\, \sum_{n=-\infty}^{\infty}f_{n}\,\frac{(U_{\rm{p}}-nw)}{E_{\rm{q}}+E_{\rm{B}}+U_{\rm{p}}-nw-\imath \eta}\,\nonumber \\
 &\times & e^{\imath (E_{\rm{q}}+E_{\rm{B}}+U_{\rm{p}}-nw)t}
\end{eqnarray}
 Substituting Eq.\ (18) for $\Gamma$ into Eq.\ (14) we obtain for $S^{(\rm{1})}_{\rm{fi}}$ the following expression
\begin{eqnarray}
S^{(\rm{1})}_{\rm{fi}} & = & \imath \int d\vec{q}\, \langle \vec{q}\mid \phi_{\rm{i}}\rangle\, \langle \vec{k}\mid V_{\rm{A}}\mid \vec{q}\rangle\, \int_{-\infty}^{\infty}dt\, e^{\imath\left(E_{\rm{k}}-E_{\rm{q}}\right)t}\nonumber \\ & \times & \sum_{n=-\infty}^{\infty}f_{n}\,\frac{(U_{\rm{p}}-nw)\, e^{\imath (E_{q}-E_{\rm{i}}+U_{\rm{p}}-nw)t}}{E_{\rm{q}}+E_{\rm{B}}+U_{\rm{p}}-nw-\imath \eta}
\end{eqnarray}
the temporal integral gives $2\pi\delta(E_{\rm{k}}+E_{\rm{B}}+U_{\rm{p}}-n\omega)$ and therefore we finally obtain
\begin{equation}
S^{(\rm{1})}_{\rm{fi}}=2\pi\imath \sum_{n=n_{{0}}}^{\infty}\delta (E_{k}+E_{\rm{B}}+U_{{p}}-n\omega)\,
 \int d{\bf q}\, \langle {\bf q}\mid \phi_{\rm{i}}\rangle\, \langle {\bf k}\mid V_{\rm{A}}\mid {\bf q}\rangle
  \frac{U_{{p}}-n\omega}{E_{q}-E_{\rm{k}}-\imath \eta}
 \, f_{n}
 \end{equation}
which, according to Eq.\ (10) in the supplementary material, is exactly the singularity in  $m=n$ term in Eq.\ (6); i.e., $S^{{(1)}}_{\rm{ir}}$. The fundamental point here which is of great importance is that the singularity is identified to be due to Coulomb forward scattering in the vicinity of the atomic core (since $\balpha=0$) and this identification led to the isolation and hence the removal of the singularity.

\bibliography{basename of .bib file}